# Relic Neutrinos, Z–Bursts, and Cosmic Rays above $10^{20}$ eV

Thomas J Weiler†[1]

† Department of Physics & Astronomy, Vanderbilt University, Nashville, TN 37235, USA (email: tom.weiler@vanderbilt.edu)

**Abstract.** The observation of cosmic–ray events above the Greisen–Kuzmin–Zatsepin (GZK) cutoff of $E_{\rm GZK} \sim 5 \times 10^{19}$ eV challenges orthodox modeling. We discuss a possible solution which uses standard hot Big Bang cosmology and Standard Model (SM) particle physics augmented only by $\lesssim$ eV neutrino masses as suggested by solar, atmospheric, and terrestrial neutrino detection. In this scheme, cosmic ray neutrinos from distant, highest–energy sources annihilate resonantly on the relic–neutrino background to produce $Z$–bosons. The highly–boosted ($\gamma_Z \sim 10^{11}$) $Z$'s instantly decay, producing "Z-bursts" of highly–collimated jets of hadrons and photons. The burst content includes, on average, twenty photons and two nucleons with super–GZK energies. We show that the probability for each neutrino with energy within a fraction $\Gamma_Z/M_Z$ of the resonant value $E_R = 4\,({\rm eV}/m_\nu) \times 10^{21}$ eV to annihilate within the halo of our galactic supercluster is likely within an order of magnitude of 1%. Depending on the magnitude of the cosmic neutrino flux above $10^{20}$ eV, this "local" rate for primary production may be high enough to produce the cosmic ray events observed above the GZK cutoff. Several tests of this $Z$–burst hypothesis for generating super–GZK events are presented, including (i) a new cutoff energy at $E_R$; (ii) a large $\gamma/p$ ratio for primaries near the upper end of the spectrum; (iii) directional pairing of events, and pointing to their cosmic sources; and
(iv) a neutrino flux above the GZK cutoff energy which is possibly measurable directly in proposed $\gtrsim 10^{11}$ ton cosmic–ray detectors.

## 1. Introduction: the cosmic ray puzzle above $10^{20}$ eV

It has long been anticipated that the highest–energy cosmic–ray primaries would be protons from outside the galaxy, perhaps produced in active galactic nuclei (AGNs). Since the mid–sixties it was also anticipated that the highest energies for protons arriving at earth would be $E_{\rm GZK} \sim 5 \times 10^{19}$ eV. The origin of this Greisen–Kuzmin–Zatsepin (GZK) cutoff [1] is the degradation of the proton energy by the resonant scattering process $p + \gamma_{2.7K} \to \Delta^* \to N + \pi$ when the proton is above the resonant threshold for $\Delta^*$ production; $\gamma_{2.7K}$ denotes a photon in the $2.7K$ cosmic background radiation. A proton produced at its cosmic source at a distance $D$ from earth with an initial energy $E_p$ well above $E_{\rm GZK}$ will on average lose 20% of its energy per interaction length of 6 Mpc to arrive at earth with only a fraction $\sim (0.8)^{D/6\,{\rm Mpc}}$ of its original energy. Therefore, proton energy is not lost significantly only if the highest–energy protons come from relatively nearby sources, $\stackrel{<}{\sim}$ 50 to 100 Mpc [2]. However, no AGN sources are known to exist within 100 Mpc of earth.[2] Any observation of air–shower events above $5 \times 10^{19}$ eV challenges the theory. To date, twenty–four air–shower events with energies near and above $10^{20}$ eV have been observed by the Volcano Ranch, Haverah Park, Yakutsk, Fly's Eye, Akeno/AGASA, and most recently, HiRes collaborations [4]. It is now clear that the highest–energy cosmic–ray spectrum extends beyond the GZK cutoff.

A primary nucleus mitigates the cutoff problem (energy per nucleon is reduced by 1/A), but it has additional problems: above $\sim 10^{19}$ eV nuclei may be photo–dissociated by the 2.7K background, and possibly disintegrated by the particle density ambient at the astrophysical source. Gamma–rays and neutrinos are other possible primary candidates for the highest–energy events. The gamma–ray possibility is not supported by the time–development of the Fly's Eye event, but is not ruled out for this event. However, the mean free path for a $\sim 10^{20}$ eV photon to annihilate to $e^+e^-$ on the radio background is believed to be $\stackrel{<}{\sim}$ 10 Mpc based on recent estimates of the background, and the density profile of the highest–energy Yakutsk event showed a large number of muons which may argue against gamma–ray initiation.

### 1.1. Paired–events and neutrino primaries above $10^{20}$ eV

Turning to the possibility that the primaries may be neutrinos, one encounters an immediate obstacle: the Fly's Eye event occured high in the atmosphere, whereas the expected event rate for early development of a neutrino–induced air–shower is down from that of an electromagnetic or hadronic interaction by six orders of magnitude. On the other hand, the neutrino hypothesis is supported by evidence [5] that some of the highest–energy primaries have common arrival directions, with arrival times displaced by only $\sim 2$ to 3 years. Of the 47 AGASA events above $4 \times 10^{19}$ eV, 9 are contained in three doublets and one triplet with separation angle less than the angular resolution of $2.5°$. The chance probability of this clustering occuring in an isotropic distribution is less than 1%. The chance probability for the triplet alone is only 5%. Of the seven events above $10^{20}$ eV, three are counted among the doublet events. Such event–pairing argues for a common source of some duration, emitting stable neutral primaries having a small magnetic moment. The stability requirement is that $c\tau$ exceed $D \sim D_H$, where the Hubble distance is $D_H \equiv c\, H_0^{-1} = 1.4\, h_{65}^{-1} \times 10^{28}$ cm, and $h_{65}$ is the Hubble parameter in units of 65 km/s/Mpc. The resulting limit on the rest frame lifetime is then $\tau_{\rm RF} \gg (m/E_{\rm GZK})(D_H/c) = 10^{-2}(m/{\rm eV})h_{65}^{-1}s$; for a GeV primary, this is about one year, while for an eV neutrino it is as short as tens of milliseconds. The requirement that the primary be neutral is necessary to allow straight trajectories for the paired primaries as they travel through the (probably nanogauss or greater) extragalactic magnetic fields. The requirement that the primary have a small magnetic moment is necessary to avoid significant energy losses via magnetic dipole interactions with the ambient 2.7K microwave background. Neutrino primaries satisfy these three criteria.

---

[2] The suggestion has been made that hot spots of radio galaxies in the supergalactic plane (e.g. M87) at distances of tens of megaparsecs may be the sources of the super–GZK primaries [3]. Statistical support for this hypothesis is weak at present.

Let us call the distance over which a stable particle can propagate without losing more than an order of magnitude of its energy the GZK distance, $D_{\rm GZK}$. For a photon it is a few Mpc at $10^{20}$ eV, rising to $\mathcal{O}(100)$ Mpc at $10^{22}$ eV, with the exact numbers depending on the strengths of the diffuse radio background. For a proton it is $D_{\rm GZK} \sim 50$ to 100 Mpc. Here we propose that the primary particles which propagate across cosmic distances above the GZK cutoff energy are neutrinos, which then annihilate with relic neutrinos within the GZK zone $(D < D_{\rm GZK})$ to create a "local" flux of nucleons and photons above $E_{\rm GZK}$ [6, 7]. The annihilation rate depends upon the relic neutrino background reliably predicted by Big Bang cosmology, neutrino clustering in gravitational potentials, and the Standard Model (SM) of particle physics augmented with neutrino mass values suggested by recent oscillation data. For a sufficient cosmic neutrino flux, the hypothesis successfully explains the observed air–showers above $E_{\rm GZK}$.

## 2. $Z$–bursts from resonant neutrino annihilation

It was noted many years ago [8] that the mean free path (mfp), $\lambda_j = [n_{\nu_j}\, \sigma_{ann}(\nu_j + \bar{\nu}_j \to Z)]^{-1}$, for a cosmic ray neutrino to annihilate at the $Z$ resonance on a background of nonrelativistic relic antineutrinos (and vice versa) having mass $m_j$ and density $n_{\nu_j}$ is only slightly larger than the Hubble size of the Universe, $D_H = 1.4\, h_{65}^{-1} \times 10^{28}$ cm. This means that the annihilation probability per cosmic distance of travel may be significant and that absorption spectroscopy is a possible means, in principle, for determining neutrino masses [8]. The energy of the neutrino annihilating at the peak of the $Z$–pole is

$$E_{\nu_j}^R = M_Z^2/2m_{\nu_j} = 4\,(\text{eV}/m_{\nu_j}) \times 10^{21}\,\text{eV}\,. \qquad (1)$$

At a given resonant energy $E_{\nu_j}^R$, only relic neutrinos with the $j^{th}$ mass $m_j$ may annihilate. (We will sometimes use $E_R$ generically for resonant energy with the understanding that there are really three different resonant energies, one for each neutrino mass.)

The invariant energy–averaged annihilation cross section for the process $\nu_j + \bar{\nu}_j \to Z$ is given by the integral over the $Z$ pole. In the SM this is

$$\langle \sigma_{\text{ann}} \rangle \equiv \int \frac{ds}{M_Z^2} \sigma_{ann}(s) = 4\pi G_F/\sqrt{2} = 4.2 \times 10^{-32}\text{cm}^2, \qquad (2)$$

with $s$ the square of the energy in the center of momentum frame. When neutrino mixing–angles are introduced in the augmented SM, this result continues to hold for each neutrino type $j$ in the flavor or mass basis, since the annihilation mechanism is a neutral current process. The energy–averaged annihilation cross section $\langle \sigma_{\text{ann}} \rangle$ is the effective cross section for all neutrinos within $\frac{1}{2}\delta E_R/E_R = \Gamma_Z/M_Z = 3\%$ of their peak annihilation energy. We refer to neutrinos with resonant flavor $j$ and with energy in the resonant range $0.97\, E_{\nu_j}^R$ to $1.03\, E_{\nu_j}^R$ as "resonant neutrinos."

Each resonant neutrino annihilation produces a $Z$ boson which immediately decays (its lifetime is $3 \times 10^{-25}$ s in its rest frame). 70% of these decays are hadronic, consisting of a particle burst known to include on average about one baryon–antibaryon pair, seventeen charged pions, and ten neutral pions [9]. The ten $\pi^0$'s decay to produce twenty high–energy photons. We refer to the end product of this $Z$ production and hadronic decay as a "$Z$–burst." The nucleons (we now mean "nucleons" to include the antinucleons as well) and photons in the $Z$–burst are candidates for the primary particles with energy above the GZK–cutoff [6, 7]. The two nucleons may be protons or neutrons; if the neutron decays, it simply produces a proton with nearly the same energy as the parent neutron.

If the $Z$–burst points in the direction of earth and occurs within the GZK distance, then one or more of the photons and nucleons in the burst may initiate a super–GZK air–shower at earth. The mean multiplicity in $Z$ decay is about 30 [9]. This dilutes the energy per hadron somewhat compared to $E_{\nu_j}^R$, but it also provides a larger flux per burst. Shown in Figure 1 is a schematic of the $Z$–burst mechanism within the GZK zone.

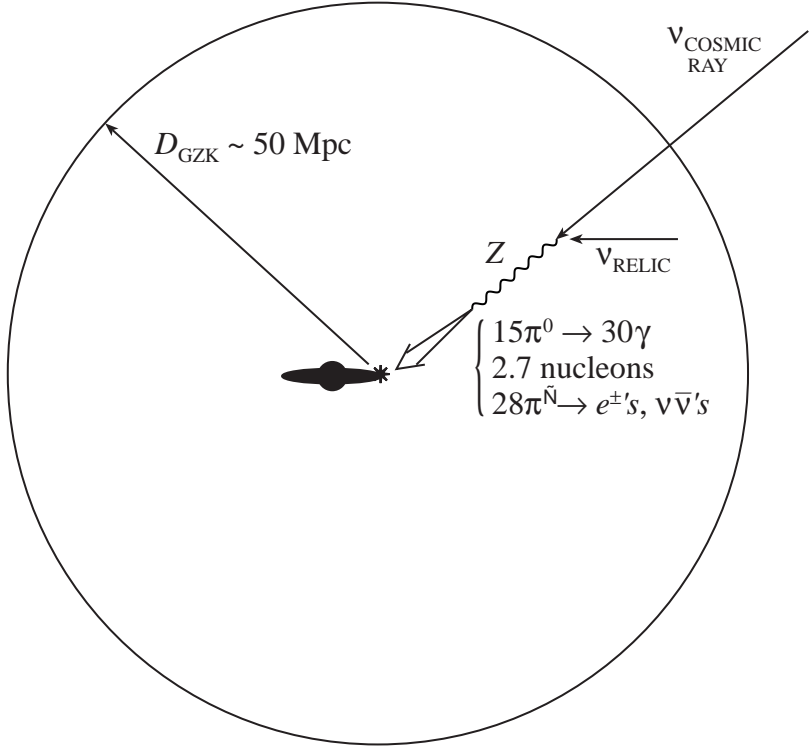

Figure 1. Schematic diagram showing the production of a $Z$–burst resulting from the resonant annihilation of a cosmic–ray neutrino on a relic (anti)neutrino. If the $Z$–burst occurs within the GZK zone ($\sim$ 50 to 100 Mpc) and is directed towards the earth, then photons and nucleons with energy above the GZK cutoff may arrive at earth and initiate super–GZK air–showers.

*2.1. Neutrino mass and resonant energy*

Note that two crucial elements are required for this mechanism to produce super–GZK air–showers: the existence of a neutrino flux at $\gtrsim 10^{21}$ eV, and the existence of a neutrino mass in the 0.1 to a few eV range. Concerning possible neutrino masses, the simplest explanation for the anomalous atmospheric–neutrino flavor–ratio [10] and its zenith–angle dependence [11] is neutrino oscillations driven by a mass–squared difference of $\delta m^2_{\rm atm} \sim 10^{-3}$ to $10^{-2}$ eV$^2$ [12], which implies a neutrino mass of *at least* 0.03 to 0.1 eV. Also, the recent LSND measurement appears to indicate a mass–squared difference of $\delta m^2_{\rm LSND} \sim 0.2$ to 4 eV$^2$ [13], from which one deduces a neutrino mass of at least 0.5 to 2 eV.

Upper bounds on the neutrino masses are available from tritium decay experiments and from large–scale structure formation studies. From the end point spectrum in tritium decay, one infers the upper bound $m_\beta < 4.4$ eV for the weighted rms mass $m_\beta \equiv (\sum_j |U_{ej}|^2 m_j^2)^{\frac{1}{2}}$. In a three neutrino universe, the smallness of the neutrino mass–splittings deduced from atmospheric and solar oscillations ($\delta m < \sqrt{\delta m^2_{\rm atm}} < 0.1$ eV) leads to the universal bound $m_{\nu_j} < 4.4$ eV for all three neutrino masses [14]. In a four neutrino universe adopted to accommodate the LSND oscillation signal, the analogous bound on each of the four neutrino masses is
$m_{\nu_j} < [(4.4)^2 + \delta m^2_{\rm LSND}]^{\frac{1}{2}} \sim 5.4$ eV [14].

The nonobservation of neutrinoless double-beta decay provides the further upper bound $|\sum_j U_{ej}^2 m_j| < 0.2$ eV for Majorana neutrino masses [15]. This bound is restrictive in particular models where the complex mixing elements $U_{ej}$ are known. In general, however, cancellations may occur in the sum, in which case the bound is compatible with large neutrino masses.

According to Big Bang cosmology, the fraction of closure density provided by the masses of nonrelativistic neutrinos is
$\Omega_\nu = 0.025\, h_{65}^{-2} \sum_j (m_{\nu_j}/{\rm eV})$. One sees that eV neutrino masses are required if neutrino hot dark matter is to contribute in any significant way to the evolution of large–scale structure [16]. Studies of structure formation suggest that $\Omega_\nu \lesssim 0.15$ [17], in which case no single neutrino mass may exceed $\sim 6$ eV. Particular studies of cosmic structure formation give upper limits to $m_\nu$ in the 2 to 6 eV range [18]. This is consistent with, and complementary to, the tritium bound just discussed.

According to eqn. (1), an upper bound on the neutrino mass of 4 eV implies that no $Z$–burst energy will fall below $10^{21}$ eV. This is a comfortable factor of 20 or so above the GZK cutoff. From the lower bounds on neutrino mass deduced from the SuperK and LSND experiments, one gets upper bounds on the $Z$–burst energy of $10^{23}$ and $10^{22}$ eV, respectively. It is uncanny that the allowed range for $Z$–burst energies of $10^{21}$ to $10^{23}$ eV, deduced from cosmology, tritium decay, and oscillation experiments, is just right for extending the air-shower spectrum an order of magnitude or two above the GZK cutoff!

*2.2. Relic neutrino number density.*

The annihilation/$Z$–burst rate depends on the relic–neutrino number–density. Hot Big Bang cosmology predicts the *mean* neutrino density of the universe. The density of neutrinos with mass below the neutrino decoupling temperature $\sim 1$ MeV is given by a relativistic Fermi-Dirac distribution, characterized by the single temperature parameter $T_\nu \sim 1.95$K. $T_\nu$ is a factor of $(4/11)^{1/3}$ less than that of the photon temperature $T_\gamma = 2.73$K as a result of photon reheating from the era of $e^+e^- \to \gamma\gamma$ annihilation. As it is the momentum (not the energy) which red–shifts with the temperature, the mean momentum of the neutrinos today is $p_\nu \sim 3\,T_\nu \sim 6 \times 10^{-4}$ eV. Consequently, relic neutrinos with a mass exceeding this value have an energy dominated by their rest mass. The resulting mean neutrino number density is

$$\langle n_{\nu_j} \rangle = (3\,\zeta(3)/4\pi^2)\,T_\nu^3 = 54\,\text{cm}^{-3} \quad (3)$$

for each light flavor $j$, with an equal number density for each antineutrino flavor. The value of $\zeta(3)$ is $1.202\ldots$. We make two remarks here. The first is that the universe should possess an abundance of antineutrinos as well as neutrinos. The reason is that the neutrinos and antineutrinos fell out of thermal equilibrium while still relativistic (hence the name "hot dark matter"). This is in contrast to the electron, proton and neutron, for which the nonrelativistic Boltzmann factor exponentially suppressed the antiparticle. The second remark is that the predicted relic–neutrino density is normalized to the relic photon density, which is measured, via the temperature relation $T_\nu^3 = \frac{4}{11}T_\gamma^3$. Consequently, the predicted mean density of $\langle n_{\nu_j} \rangle = 54\,\text{cm}^{-3}$ in the absence of a neutrino chemical potential must be considered firm. The possibility of a nonzero chemical potential and its implication for the relic density and annihilation rate are discussed in section 2.6.

*2.3. Probability for resonant annihilation to a $Z$–burst*

In a universe where the neutrinos are nonrelativistic and uniformly distributed, the mean annihilation length for neutrinos at their resonant energy would be

$$\lambda_j = (\langle\sigma_{\text{ann}}\rangle\langle n_{\nu_j}\rangle)^{-1} = 4.4 \times 10^{29}\text{cm}, \quad (4)$$

which is $30\,h_{65}$ times the Hubble distance [8]. A cosmic ray neutrino arriving at earth from a cosmically distant source will have traversed approximately a Hubble distance of space, so its annihilation probability on the relic–neutrino sea is roughly $D_H/\lambda_j = 3.0\, h_{65}^{-1}\,\%$ (neglecting cosmic expansion).

For a more careful derivation, we let $F_{\nu_j}(E_\nu, x)$ denote the flux of the $j^{\text{th}}$ neutrino flavor, as would be measured at a distance $x$ from the source, with energy within $dE$ of $E_\nu$. The units of $F_{\nu_j}(E_\nu, x)$ are neutrinos/energy/area/time/solid angle. This flux may be quasi–isotropic ("diffuse"), as might arise from a sum over cosmically–distant sources such as AGNs; or it may be highly directional, perhaps pointing back to sources within our supergalactic plane. The production rate of $Z$'s with energy within $dE$ of $E_\nu$, per unit length, area, and solid angle, is therefore $dF_Z(E_\nu, x)/dx = \sigma_{ann}(E_\nu) n_{\nu_j}(x) F_{\nu_j}(E_\nu, x)$. Integrating this equation over the distance $D$ from the emission site to earth, and integrating over neutrino energy then gives the total rate of resonant annihilation, *i.e.* $Z$–burst production (in the narrow resonance approximation), within the distance $D$:

$$F_Z(D) = \delta F_{\nu_j}(D) = E_R F_{\nu_j}(E_R, 0) \int \frac{ds}{M_Z^2} \left[ 1 - \exp\{-\sigma_{ann}(s)\, S_j(D)\} \right], \tag{5}$$

where $S_j(D) \equiv \int_0^D dx\, n_{\nu_j}(x)$ is the neutrino column density from earth to the source at distance $D$. If $\sigma_{ann}(s) S_j(D)$ is small compared to one, then

$$F_Z(D) \approx E_R\, F_{\nu_j}(E_R, 0)\, \langle \sigma_{\text{ann}} \rangle S_j(D). \tag{6}$$

For $S(D) \ll 1/\sigma_{ann}(s)$, the rate for $Z$–burst generation depends linearly on the relic–neutrino column density. Writing $S(D)$ as $\langle n_{\nu_j} \rangle \times D$, we make contact with our previous simple estimate, namely $F_Z(D)/E_R F_{\nu_j}(E_R, 0) = D/\lambda_j$, with $\lambda_j$ given in eqn. (4). Also for $S(D) \ll 1/\sigma_{ann}(s)$, the attenuation of the neutrino flux over the distance from the source to the GZK zone will be small, and we may set $D$ in eqn. (6) equal to the GZK distance to get the $Z$–burst rate within the GZK zone.

For each 50 Mpc of travel through the mean neutrino density, the probability for a neutrino with resonant energy to annihilate to a $Z$–boson is $3.6 \times 10^{-4}$. Since the branching fraction for a $Z$ to decay to hadrons is 70%, one part in 4000 of the resonant neutrino flux will be converted into a $Z$–burst containing ultrahigh–energy photons and nucleons within $D_{\text{GZK}} \sim 50$ Mpc of earth in this homogeneous, zero chemical potential neutrino universe.

However, we live in a matter–rich portion of the universe. The Supergalactic Cluster, the Local Group, and the Galactic Halo each provides a "local potential well" which is expected to trap neutrinos and enhance the local relic–neutrino density compared to the universal average. Moreover, any neutrino–antineutrino asymmetry (parameterized by a chemical potential) increases the sum of the neutrino plus antineutrino relic densities. Therefore, the value of 0.025% for the probability of a resonant neutrino creating a $Z$–burst within the GZK zone, obtained without neutrino clustering or asymmetry, is the <u>absolute</u> <u>minimum</u> annihilation probability in a Big Bang universe. With clustering and/or asymmetry, the probability will be larger.

*2.4. Rate enhancement from neutrino clustering.*

Gravitational attraction will cause neutrinos to cluster somewhat in the potential wells of baryonic matter. Let us define some fiducial values for cluster sizes ($L_s$) and the mean separation distance of the clusters ($L_{ss}$). We take $D_S = 20$ Mpc for the diameter of the Virgo Supercluster, the only supercluster well within our GZK zone, and 100 Mpc, typical of the distance across a cosmic void, for the supercluster mean separation distance $D_{SS}$. We take $D_C = 5$ Mpc for the typical diameter of galactic clusters and $D_{CC} = 50$ Mpc as the typical distance between neighboring galactic clusters. For our Local Group of $\sim 20$ galaxies, we take $D_{Gp} = 2$ Mpc, and $D_{Gp-Gp} = 20$ Mpc for the mean separation distance between groups. $D_G = 50$ kpc is a typical diameter of galactic halos (including our own) and $D_{GG} = 1$ Mpc is a typical distance between neighboring galaxies. With these fiducial values for sizes and separation distances, the respective density enhancements $(L_{ss}/L_s)^3$ are of order 100, $10^3$, $10^3$, and $10^4$, for the Supercluster, Cluster, Local Group, and Galactic halo, assuming that neutrino clustering is more or less as efficient as baryonic clustering for these scales.

The geometrical scaling law that governs the increase in annihilation rate when neutrinos are clustered rather than distributed uniformly throughout the universe is $(L_{ss}/L_s)^3 (L_s/D_H) = L_{ss}^3/L_s^2 D_H$. If the neutrino clustering scale $L_s$ is less than the GZK distance of 50 to 100 Mpc, then the $Z$–burst production probability within the GZK zone also scales with this factor, yielding $(L_{ss}^3/D_H L_s^2)(3.0\, h_{65}^{-1}\,\%)$. Including the 70% hadronic branching ratio of the $Z$, one then gets the probabilities 1.3%, 2.4%, 1.0%, and 0.2% (independent of $h_{65}$), for $Z$–burst production by a neutrino of relevant flavor at resonant energy traversing the Supercluster, Cluster, Local Group, and Galactic halo, respectively. The larger clusters all give a robust probability, within a factor of two of each other, and of order of a per cent. This is our main result [6], which we repeat for emphasis: *the probability for cosmic ray neutrinos at their resonant energy to annihilate within the $\sim 50$ Mpc zone of earth is likely within an order of magnitude of 1%, with the exact value depending on unknown aspects of neutrino mixing and relic neutrino clustering.*

If the neutrino cluster is not isotropic with respect to our preferred position at earth, the $Z$–burst rate will not be isotropic either. The anisotropic rate is easily accounted for by generalizing the distance $D$ from the $Z$–burst to earth to be a vector $\hat{D}$. The column density integral encountered earlier then becomes an integral of $n_{\nu_j}(\hat{x})$ along the vector $\hat{D}$. There is weak evidence that the super–GZK events may correlate with the Supergalactic plane [3]. Such a correlation would arise naturally in the model presented here if the SG plane provided either the super–GZK neutrino flux or the potential well in which neutrinos are clustered (or both).

We now must discuss the efficiency of neutrino clustering. One expects the relic neutrinos to be less clustered than the baryons, especially on scales as small as the Galactic halo, for several reasons. First of all, neutrinos do not dissipate energy as easily as electrically charged protons do. Secondly, neutrinos have a much larger Jeans ("free–streaming") length than do baryons at the crucial time when galaxies start to grow nonlinearly. And thirdly, Pauli blocking presents a significant barrier to clustering of light–mass fermions. As a crude estimate of Pauli blocking, one may use the zero temperature Fermi gas as a model of the gravitationally bound halo neutrinos. Requiring that the Fermi momentum of the neutrinos not exceed the virial velocity $\sigma \sim \sqrt{MG/L}$ within the cluster, one gets the phase space constraint [19]

$$n_{\nu_j}/54\,\mathrm{cm}^{-3} \stackrel{<}{\sim} 10^3 (m_{\nu_j}/\mathrm{eV})^3 (\sigma/200\mathrm{kms}^{-1})^3. \qquad (7)$$

The same constraint holds for $n_{\bar{\nu}_j}$. The virial velocity within our Galaxy is a couple hundred km/sec, whereas virial velocities for rich galactic clusters are a thousand km/s or more. Thus it appears that Pauli blocking allows significant clustering on the Galactic scale only if $m_\nu \stackrel{>}{\sim} 1$ eV, but allows clustering on the larger scales for $m_\nu \stackrel{>}{\sim} 0.1$ eV. The free–streaming argument also favors clustering on the larger scales. This is just as well, for we have shown that it is the larger scales of clustering that give the $\mathcal{O}(1\%)$ probability for annihilation to a $Z$–burst.

*2.5. Absorption versus emission mass spectroscopy for neutrinos.*

The original resonant annihilation proposal [8] focussed on absorption spectroscopy, i.e. a measurement of dips in the high energy neutrino spectrum. Measurement of the absorption energies would determine the neutrino masses via $m_j = M_Z^2 \langle N \rangle /2E$, where $\langle N \rangle \sim 30$ is the mean multiplicity in the $Z$–burst and $E$ is an observed air–shower energy. High statistics are required to measure absorption dips, for an experiment must map out a range of the spectrum. On the other hand, in the absence of background events, high statistics are not required in emission spectroscopy. In emission spectroscopy, the emitted decay products of the absorbed neutrino are measured directly. The $Z$–burst hypothesis is that the primaries which initiate the showers observed above $E_{\mathrm{GZK}}$ are the emission products of the neutrino annihilation process. Furthermore, without the bursts, the GZK cutoff prohibits any background events. Thus, by collecting events above the GZK cutoff, emission spectroscopy is realized on an event by event basis.

A short annihilation mfp ($\lambda_j$) is advantageous for an absorption spectroscopic experiment, in that it enhances the depth of the dips. However, for emission spectroscopy, where the detected particles must arise within the GZK zone, a mfp of order of $D_H$ is optimal — a longer mfp reduces the annihilation rate whereas a shorter mfp reduces exponentially the neutrino flux arriving in the GZK zone. Quantitatively, the probability for a $Z$–burst per resonant neutrino is the product of the probability for a neutrino to propagate across a distance $\sim D_H$, times the probability to annihilate within the GZK distance, i.e. $P(Z\text{–burst})=(e^{-D_H/\lambda_j})(\frac{D_{\text{GZK}}}{\lambda_j})$, for $D_{\text{GZK}} \ll \lambda_j$. This burst probability may be written $(\frac{D_{\text{GZK}}}{D_H})(re^{-r})$, where $r \equiv \frac{D_H}{\lambda_j}$. This probability is maximized at $\frac{1}{e}\frac{D_{\text{GZK}}}{D_H}$ when $r = 1$, i.e. when $\lambda_j = D_H$. This derivation assumes a static universe. The result is altered by the expansion of the universe.

*2.6. Rate enhancement from a neutrino–antineutrino asymmetry.*

A further enhancement of the neutrino density arises if the universe possesses a net neutrino number. Baryon number and lepton number are not conserved quantities in the SM, due to coherent processes at the electroweak symmetry–breaking scale. Baryon number nonconservation, along with CP–violation and out of thermal equilibrium dynamics, may explain the observed baryon asymmetry in the universe, $\Delta B \equiv (n_B - n_{\bar{B}})/n_\gamma \sim 10^{-9}$. Since the baryons are primarily protons, charge neutrality requires a similar asymmetry for the electron–positron system. The neutrino asymmetry is unknown, and poorly bounded by observation. In terms of the degeneracy parameter $\xi \equiv \mu/T$, the neutrino asymmetry is

$$\Delta \nu \equiv (n_{\nu_j} - n_{\bar{\nu}_j})/n_\gamma = 0.025(\pi^2 \xi + \xi^3). \tag{8}$$

With a large degeneracy parameter, the thermal equilibrium distributions are such that the density of one species ($n_{\nu_j}$ or $n_{\bar{\nu}_j}$) is suppressed exponentially in $\xi$ while the density of the other is enhanced as a power–law in $\xi$. Also, with $\xi \neq 0$, the sum $n_{\nu_j} + n_{\bar{\nu}_j}$ is always enhanced relative to the symmetric $\xi = 0$ sum. Studies of galaxy formation and primordial nucleosynthesis lead to the bounds [20] $-0.06 \lesssim \xi_{\nu_e} \lesssim 1.1$ and $|\xi_{\nu_{\mu,\tau}}| \lesssim 6.9$. These bounds on $\xi$ translate into $\Delta \nu_e \lesssim 0.3$, which is modest, and $\Delta \nu_{\mu,\tau} \lesssim 10$, which is large. The latter bound allows an enhancement in $n_{\nu_{\mu,\tau}}$ or $n_{\bar{\nu}_{\mu,\tau}}$ by a factor up to 80. Of course, $\xi$ cannot be so large that $\Omega_\nu$ exceeds the bound deduced from large–scale structure formation. Since $\xi^3 (m_\nu/\text{eV}) = 430\, \Omega_\nu h_{65}^2$ is valid for $m_\nu/T_\nu \gg \xi \gg 1$, we have, from $\Omega_\nu h_{65}^2 \lesssim 0.15$, that $\xi^3 (m_\nu/\text{eV}) \lesssim 65$.

It has been noted in [21] that a degeneracy parameter of $\xi \sim 5$ (giving a neutrino asymmetry factor of $\sim 4$) effects an enhancement in the neutrino or antineutrino number density of $\sim 30$, thereby decreasing the resonant annihilation mfp for antineutrinos or neutrinos to $\sim D_H$ and maximizing the $Z$–burst probability within the GZK zone to $\frac{1}{2} \times \frac{1}{e} \frac{D_{\text{GZK}}}{D_H} \sim 0.2\, h_{65}\%$. With neutrino clustering in addition to the asymmetry discussed here, this probability is larger still. The factor of $\frac{1}{2}$ in the probability reflects the fact that with a large asymmetry, one of $n_\nu$ or $n_{\bar\nu}$ is enhanced but the other is driven near zero. Consequently, the annihilation probability for the cosmic–ray $\bar\nu$ or $\nu$, respectively, is negligible.

If there is a net neutrino number enhancement of $\sim 30$, then the cosmological bound $\Omega_\nu \lesssim 0.15$ requires that $m_\nu \lesssim 0.4$ eV for the enhanced species. This in turn puts a lower bound on the resonant energy of $E_R \gtrsim 10^{22}$ eV.

We remark that although the two spin states of Majorana neutrinos are not particle and antiparticle, the same thermodynamic counting that applies to Dirac neutrinos applies also to Majorana neutrinos. Consequently, the asymmetry and density enhancement that result from the calculation apply to both kinds of neutrino. The calculation is common because the two Majorana spin states can annihilate only with each other when relativistic, like particle and antiparticle, owing to the helicity conserving nature of the vector and axial vector weak interaction in the relativistic limit. Once a density asymmetry or enhancement is fixed at the time of decoupling, when the neutrinos are relativistic, it cannot be changed (in comoving coordinates) during free expansion.

## 2.7. Two subtleties

We now discuss two subtleties that will affect the annihilation rate. The first is the depolarization of the relic neutrinos as they evolve from their relativistic state at decoupling to the nonrelativistic state which they occupy today. The second is flavor–mixing among massive neutrinos.

### 2.7.1. Neutrino depolarization.
The red–shifting of the neutrino momenta due to the expanding universe, from relativistic at decoupling to nonrelativistic today, renders the neutrinos unpolarized. As a result, if the neutrino is a Dirac particle, then the densities of the sterile right–handed neutrino and the sterile left–handed antineutrino fields are equal to the densities of the two active fields. Therefore, for Dirac neutrinos the active densities available for annihilation with the incident cosmic–ray neutrino are half of the total densities, and the $Z$–burst production probability is half of what we quote in this article. For Majorana neutrinos, there are no sterile fields and the total densities are active in annihilation.

Depolarization of Majorana neutrinos also means that $n_{\nu_L} = n_{\nu_R}$ today. Although the concept of "neutrino asymmetry" or "net lepton number" loses its meaning for Majorana neutrinos, any density enhancement due to an asymmetry at the time of decoupling remains. Majorana neutrinos are favored over Dirac neutrinos in currently popular theoretical models with nonzero neutrino mass [22].

*2.7.2. Flavor mixing in the mass basis.* Oscillation experiments strongly suggest that the mass and interaction ("flavor") bases of neutrinos do not coincide. In general, the flavor states are unitary mixtures of the mass states. Letting $\alpha = e, \mu, \tau$ label flavor and $j = 1, 2, 3$ label mass, one has $|\nu_\alpha> = \sum_j U_{\alpha j}|\nu_j>$, where $U$ is the unitary mixing matrix. Then each neutrino flavor at the resonant energy of a given mass state has a nonzero probability to annihilate, but with an extra probability factor of $|U_{\alpha j}|^2$. For example, the $\nu_\mu$'s and $\nu_e$'s from pion and mu decay will annihilate at the resonant energy of $m_2$ with the probability factors $|U_{\mu 2}|^2$ and $|U_{e2}|^2$, respectively, times what we have calculated above. The mixing factors $|U_{\alpha j}|^2$ can be easily multiplied into our result. Indications from the solar and atmospheric neutrino experiments are that the mixing factors are large; e.g., in the presently popular "bi–maximal" model [23], all the $|U_{\alpha j}|^2$ factors, except for $|U_{e3}|^2$, are in the range 0.25 to 0.5.

Flavor mixing also affects the present density of relic mass eigenstates, giving $n_{\nu_j} = \sum_\alpha |U_{\alpha j}|^2 n_{\nu_\alpha}$, where $\sum_\alpha$ is a sum over the three neutrino flavors and $n_{\nu_\alpha}$ is the $\nu_\alpha$ relic density expected in the absence of mixing. This formula results from the separation ("decoherence") of the neutrino wave function into distinct wave packets for each mass eigenstate after a Hubble–time of travel. Note that if the three $\nu_\alpha$ are the same, $\equiv n_\nu$, then unitarity gives $n_{\nu_j} = n_\nu$ for all three mass states $j = 1, 2, 3$.

Finally, we mention that the phenomenon of neutrino oscillations occurs when neutrinos are mixed. Neutrino oscillations will affect the flavor populations of the cosmic–ray neutrinos. However, they will not affect our calculation of annihilation.

## 3. Cosmic neutrino flux above $10^{20}$ eV

A considerable cosmic neutrino flux above $E_{\text{GZK}}$ is required for the $Z$–burst hypothesis to successfully explain the super–GZK events. According to eqn. (6), the requirement on the neutrino flux at the resonant energy is that the product of this flux per flavor times the resonant energy, times the annihilation probability within the GZK zone (which we have estimated to be $10^{-2\pm1}$ in section 2.4), times the photon and nucleon multiplicity per burst ($\sim 30$), is comparable to the flux of events above $E_{\text{GZK}}$; this is,

$$E_R F_{\nu_j}(E_R) \sim 10^{0.5\pm1} F_{p/\gamma}(> E_{\text{GZK}}). \tag{9}$$

The AGASA collaboration [24] shows an integrated flux above $5 \times 10^{19}$ eV of $F_{p/\gamma}(> 5 \times 10^{19}\,\text{eV}) \sim 1.0 \times 10^{-19}/\text{cm}^2/\text{s}/\text{sr}$. Based on the twenty events above $10^{20}$ eV with the best energy calibration,[3] A. Watson [25] calculates an integrated flux above $10^{20}$ eV of $F_{p/\gamma}(> 10^{20}\,\text{eV}) = 2.0 \pm 0.5 \times 10^{-20}/\text{cm}^2/\text{s}/\text{sr}$. These two fluxes are quite consistent with the spectral index of $2.8 \pm 0.3$ found for energies above $10^{19}$ eV by the AGASA collaboration [24]. It seems safe to assume that the flux $F_{p/\gamma}$ which is generated by the Z–bursts must be bracketed by these two flux values, i.e., by

$$F_{p/\gamma}(> E_{\text{GZK}}) = 10^{-19.5 \pm 0.4}/\text{cm}^2/\text{s}/\text{sr}. \tag{10}$$

This allows for some of the events just above $E_{\text{GZK}}$ to arise from relatively nearby sources via a conventional mechanism, or from statistical fluctuations or systematic errors. From eqns. (9) and (10), we arrive at the following estimate for the resonant neutrino flux:

$$E_R\, F_{\nu_j}(E_R) \sim 10^{-19 \pm 1.4}/\text{cm}^2/\text{s}/\text{sr}. \tag{11}$$

It is not unlikely that whatever mechanism produces the most energetic hadrons also produces charged pions of comparable energy. Thus, one may expect neutrino production at ultrahigh energy, coming from pion decay and subsequent muon decay. Photopion production in a low–density source with a straightforward power–law extrapolation of the proton spectrum is not promising [26]. However, a dense source such as an AGN may effectively trap hadrons so that the confined proton flux greatly exceeds a power law extrapolated from the observed cosmic ray flux [27]. The possibility of a large high–energy neutrino flux does not seem unnatural. In fact, it has been claimed that the MHD "snowplow" effect in a dense plasma is capable of producing an enhanced neutrino flux all the way up to $10^{24}$ eV, without the concomitant proton and photon fluxes violating any observational bounds [28].

There is also the possibility that the highest-energy neutrinos originate in quark jets produced in the decay of some supermassive relic particles [29] or topological defects [30], in which case the neutrino flux greatly exceeds the proton flux. However, this kind of exotic source takes us beyond SM physics, thereby mitigating the economy of the Z–burst hypothesis.

There do exist experimental upper limits on the high energy neutrino flux, derived from the nonobservation by the Fly's Eye of air-showers produced by penetrating primaries [31]. Using the QCD–corrected high–energy neutrino cross–section estimated to be $\sigma_{\nu N} \sim 8 \times 10^{-32} E_{20}^{0.4}\,\text{cm}^2$ [32], with $E_{20}$ being the neutrino energy in units of $10^{20}$ eV, I get the following for the Fly's Eye bounds: $2 \times 10^{-13}$, $3 \times 10^{-14}$, and $5 \times 10^{-16}$ per cm$^2$-s-sr at $E_\nu = 10^{17}$, $10^{18}$, and $10^{20}$ eV, respectively. Thus, the neutrino flux at $10^{17}$ eV cannot be much more than six orders of magnitude larger than the flux at the resonant energy. At present, this is a constraint on the source, not on the $Z$–burst model.

---

[3] Watson's flux is based on the twenty events above $10^{20}$ eV from Volcano Ranch (1), Haverah Park (4), Fly's Eye (1), AGASA (7), and HiRes (7); the Yakutsk events are not included here since there is some uncertainty in their energy calibration.

## 4. The p, n, $\gamma$ flux above $E_{\mathrm{GZK}}$ from $Z$–bursts

The decay products of the $Z$ are well–known from the millions of $Z$'s produced at LEP and at the SLC [9]. Among the $\sim 30$ particles in a $Z$–burst are 20 photons from $\pi^0$ decays, and 2 nucleons; these are the candidate primary particles for inducing super–GZK air–showers in the earth's atmosphere.

Without simulation and modeling it is not clear what ratio of photon to nucleon initiated air–showers is to be expected above the GZK cutoff. The *a priori* photon–to–nucleon ratio is about 10. However, the hardness of the nucleon spectrum compared to the photon spectrum as measured in $Z$–decay mitigates this ratio if a selection is made for the very highest energy particles. The mean pion energy from $Z$ decay is about a third of the mean nucleon energy [9]. The mean photon energy from $\pi^0$–decay is half again that of the pion. Complicating the predicition for the photon to proton ratio for air–shower initiation is the different physics causing the attenuation of photons on the radio background, and degradation of the proton energy on the 2.7K microwave background. Also, the development of photon–initiated air–showers above $E_{\mathrm{GZK}}$ is skewed by the coherent electromagnetic "LPM" effect and by high–altitude photon–absorption on the earth's magnetic field [33], either of which may affect the shower identification and interpretation. Finally, the average values for the multiplicities and energies presented here must be used with some caution, since fluctuations in multiplicity and particle-types per event, and in energy per individual particle, are large. For example, although the mean number of baryon–antibaryon pairs per hadronic $Z$–decay is one, the probabilities for 0, 1, 2, 3, and 4 pairs are 37%, 37%, 18%, 6%, and 1.5%, respectively, if the number of pairs is governed by Poisson statistics.

What does seem clear is that there should be a dominant fraction of photon–initiated air–showers not far below the upper end of the observed spectrum. Two recent numerical simulations of the $Z$–burst hypothesis have been performed, including updated quark–to–hadron fragmentation functions at the $Z$–resonance and all known extragalactic propagation effects. The conclusion of the first study [34] is that if the high energy neutrinos are emitted from a dense source with a spectrum extending to the resonant energy, and if relic neutrinos of $\sim$ eV mass are clustered on the Supercluster scale by a sensible amount, then the $Z$–burst mechanism can produce the events and rate observed above the GZK cutoff without violating lower energy flux limits. A further simulation result is that the fractions of photon and nucleon primaries are similar above $10^{20}$ eV, with photons dominating below $10^{20}$ eV. The second study [35] admits the possibility of a successful $Z$–burst explanation of the super–GZK events as long as extrapolation of the cosmic neutrino flux down in energy satisfies the Fly's Eye bounds discussed in section 3.

## 5. Further signatures from $Z$–bursts

The $Z$–burst hypothesis is highly predictive and testable. Here we discuss some possible signatures:
(i) As just mentioned in section 4, the $\gamma/p$ ratio for air–shower primaries should be large not too far below the end of the spectrum.
(ii) The energy of the $Z$–bursts are fixed at $4 \times (\text{eV}/m_{\nu_j}) \times 10^{21}$ eV by the neutrino mass(es). The energy of individual particles produced in the burst can approach this value but cannot exceed it. This may serve to distinguish the $Z$–burst hypothesis from some recent speculations for super–GZK events based on SUSY or GUT–scale physics, in which cutoff energies are expected to be much higher.
(iii) From the highest super–GZK event energy $E^{\max}$, one can deduce an upper bound on the neutrino mass of

$$m_\nu < M_Z^2/2E^{\max} = 4\,(10^{21}\text{eV}/E^{\max})\,\text{eV}\,. \tag{12}$$

Similarly, from the mean energy $\langle E \rangle$ of super–GZK events one can estimate the mass of the participating neutrino flavor via

$$m_\nu = M_Z^2/2E_R \sim M_Z^2/(2\langle N\rangle\langle E\rangle) \sim 1.3 \times (10^{20}\text{eV}/\langle E\rangle)\,\text{eV}\,. \tag{13}$$

If there is a selection bias toward events at higher energy, then this formula gives a lower bound on the neutrino mass.
(iv) The highest–energy neutrino cosmic–ray flux should point back to its sources of origin, and the super–GZK event arrival directions should point back to these same sources. The AGASA observation of directional pairing (section 1.1) of highest–energy events suggests that this is happening. Indeed, a possible correlation between arrival directions of the $E \gtrsim 10^{20}$ eV events and the locations of compact radio quasars has recently been noted [36].

The gamma primaries from $Z$–bursts must point straight back to their source, whereas the nucleon primaries from $Z$–bursts will be somewhat bent by the magnetic fields in the GZK zone. After a randow walk through a distance $D$ of magnetic domains with coherence length $\lambda$, the nucleon bends through angle

$$\delta\theta \sim 0.5° \times (D_{\rm Mpc}\lambda_{\rm Mpc})^{\frac{1}{2}} B_{\rm nG}\, E_{20}\,. \tag{14}$$

Here $B_{\rm nG}$ is the magnetic field in nanogauss. So, $10^{20}$ eV nucleons coming from 50 Mpc will bend a few degrees if they encounter nanogauss domains of Mpc size. Conventional wisdom is that extragalactic fields of strength nanogauss or less are to be expected [37]. However, there is a recent claim that the extragalactic fields may actually be of microgauss strength [38]. If true, then nucleons produced in $Z$–bursts will bend dramatically, and bear no directional correlation whatsoever with the cosmic sources.

Note that neutrons will bend as much as protons for two reasons. First, for each 6 Mpc (mfp for photopion production off the 2.7K background) of travel, the proton and neutron have an equal chance to become each other via charged pion emission. Secondly, for each neutron decay length $c\tau_n = E_{20}$ Mpc traveled, the neutron has a probability $1 - e^{-1} = 63\%$ to decay into a proton with negligible energy lost to the accompanying electron and neutrino.

(v) There could be a "neutrino pile–up" two decades of energy below $E_R$. These pile–up neutrinos are the result of the hadronic decay chain which includes $Z \to \sim 17\,(\pi^{\pm} \to \nu_\mu + \mu^{\pm} \to \nu_\mu + \bar\nu_\mu + \nu_e/\bar\nu_e + e^{\pm})$; *i.e.* each of the 70% of the resonant neutrino interactions which yield hadrons produces about 50 neutrinos with mean energy $\sim E_\pi/4 \sim E_R/120$. These neutrinos are in addition to the neutrinos piling up from the decay of pions photo–produced by any super–GZK nucleons scattering on the 2.7K background.

(vi) The Lorentz factor of the bursting $Z$ is

$$\gamma_Z = E_R/M_Z = M_Z/2m_\nu = 0.9\,(m_\nu/{\rm eV})^{-1} \times 10^{11}. \tag{15}$$

The $Z$–decay products which in the $Z$ rest frame lie within the forward hemisphere are boosted into a highly–collimated lab–frame cone of half–angle $1/\gamma_Z = 2(m_{\nu_j}/{\rm eV}) \times 10^{-11}$ radians. $Z$–bursts originating within $20\,({\rm eV}/m_{\nu_j})$ parsecs of earth, if directed toward the earth, arrive with a transverse spatial spread of less than one earth diameter. It is therefore possible (but unlikely) for the decay products of a single $Z$–burst to initiate multiple air–showers. A large area surface array (*e.g.* the Auger or Telescope Array projects) or an orbiting all–earth observing satellite (*e.g.* the OWL/AirWatch proposal) could search for these coincident showers.

There are a few more remarks of interest concerning the $Z$–burst hypothesis:

(i) If the $Z$–burst hypothesis is correct, then the neutrino flux at $E \sim 10^{22}$ eV is sufficiently high that a direct measurement of it is possible with a teraton ($10^{12}$ ton) detector. With $\sigma_{\nu N} \sim 8 \times 10^{-32}\,E_{20}^{0.4}\,{\rm cm}^2$ [32], and the neutrino flux given in eqn. (11), one has (1 ton = $0.6 \times 10^{30}$ nucleons):

$$\text{Events/year/sr/ton} = 10^{-13.2 \pm 1.4} \left(\frac{E_R}{10^{20}\text{eV}}\right)^{n-1} \int dE_{20}\, E_{20}^{(0.4-n)} \quad (16)$$

within the energy range given by the integration limits. Here $n$ is the spectral index of the neutrino flux in the vicinity of the resonant energy. Invoking as an example a flat spectrum ($n = 0$) above $E_{\text{GZK}}$ up to $E_{\text{max}}$, one gets $\sim 10^{-13 \pm 1.4} \left(\frac{E_{\text{max}}}{E_R}\right)\left(\frac{E_{\text{max}}}{10^{20}\text{eV}}\right)^{0.4}$ events/yr/sr/ton above the GZK cutoff. For the possible values $E_{\text{max}} \sim 10\, E_R \sim 10^{23}$eV, the rate from this flat spectrum example is $\sim 10^{-10.8 \pm 1.4}$ events/yr/sr/ton, which is probably observable in a teraton detector. Whereas the volumes of operational detectors are $\sim 10^9$ tons for AGASA and $10^{10}$ tons for HiRes, the volumes for the larger of the proposed detectors are $10^{11}$ tons for Auger and the Telescope Array, and $10^{12-13}$ tons for OWL/AirWatch. Since neutrinos with energy above a PeV have a charged–current interaction length in matter less than the earth's diameter [39], the neutrino signature of interest in these detectors is a penetrating horizontal shower.

A novel proposal for measuring the highest–energy neutrino flux monitors for radio pulses from the limb of the moon, which may be produced by penetrating high energy neutrinos [40]. The idea is to use radio to "see" the interactions of neutrinos traversing a small column–density of matter. For detectors in earth orbit, neutrino interactions in the limb of the earth could provide an analogous radio signal.

(ii) If the highest–energy neutrino flux is nearly isotropic, then the super–GZK event directions should correlate with the spatial distribution of the relic neutrino density. The solid angles subtended by any nearby halos may offer preferred directions for super–GZK events. As discussed is section 2.4, the Supergalactic plane may be the most probable cluster domain for neutrinos. Perhaps the angular distribution of super–GZK events can be used to perform neutrino–cluster tomography.

(iii) If the super-GZK events are due to neutrino annihilation on relics as hypothesized here, and if the high–energy neutrino flux is eventually measured, then an estimate of the relic–neutrino column density out to $D_{\text{GZK}} \sim 50$ to 100 Mpc may be made. Solving eqn. (6) for $S_j(D_{\text{GZK}})$ and replacing $F_Z(D_{\text{GZK}})$ with $F_{p/\gamma}(> E_{\text{GZK}})/\langle N \rangle$, we have

$$\begin{aligned} S_j(D_{\text{GZK}}) &\sim \frac{F_{p/\gamma}(> E_{\text{GZK}})}{E_R\, F_{\nu_j}(E_R)\, \langle \sigma_{\text{ann}} \rangle \langle N \rangle} \\ &= 0.8 \times 10^{30} \times \frac{F_{p/\gamma}(> E_{\text{GZK}})}{E_R F_{\nu_j}(E_R)}\, \text{cm}^{-2}. \end{aligned} \quad (17)$$

Here $\langle N \rangle$ is the mean number of nucleons and protons in the $Z$–burst which are above $E_{\text{GZK}}$. For the numerical example we have taken $\langle N \rangle \sim 30$, which applies if all the nucleons and photons in the $Z$–burst are above $E_{\text{GZK}}$. An estimate of $F_{p/\gamma}(> E_{\text{GZK}})$ based on recent data was given in eqn. (11). The feasibility of a future measurement of $E_R F_{\nu_j}(E_R)$ was discussed in section 3. (If $F_{\nu_j}(E)$ is measured below the resonant energy, an estimate of the neutrino column density can still be made by model extrapolation of the flux to $E_R$.) Thus, an experimental determination of $S_j(D_{\text{GZK}})$ seems possible in the not too distant future.

## 6. Summary and prospects

In summary, if one or more neutrino mass is within the range $\sim 0.04 - 1.0$ eV, and if there is a sufficient flux of cosmic ray neutrinos at $\gtrsim 10^{21}$ eV, then $\nu_{\text{cr}} + \bar{\nu}_{\text{relic}}$ (or vice versa) $\to Z \to$ *nucleons and photons* within the GZK volume $\sim (50\text{Mpc})^3$ of earth may be the origin of air–shower events observed above the GZK cutoff. If the hypothesis is correct, then air–shower observations may have already shown the existence of the relic–neutrino gas liberated from the primordial early–universe plasma when the universe was only one second old.

There are good prospects for more cosmic ray data soon at the highest energies. The AGASA and HiRes experiments are active. In the near future, Auger and the Telescope Array will become operational and increase the sensitivity by a factor of 10 to 100. In the more distant future, the proposed OWL/AirWatch satellite experiment, with fluorescence detectors looking down at our atmosphere, may become operational and increase the sensitivity by yet another factor of 10 to 100.

Possible signatures to validate or invalidate the $Z$–burst hypothesis presented here will be forthcoming. Signatures are abundant. Several were listed in section 5. The most striking of these are:
(i) a new cutoff energy at $E_R = 4\,(\text{eV}/m_\nu) \times 10^{21}$ eV;
(ii) a large $\gamma/p$ ratio for primaries near the upper end of the observable cosmic–ray spectrum;
(iii) pairing of events in direction, and directional pointing of shower–axes to their cosmic sources;
and (iv) a neutrino flux sufficiently large above the GZK cutoff energy that direct measurement is possible with the proposed Auger, Telescope Array, and OWL/AirWatch detectors.

## Acknowledgements

This work was supported in part by the U.S. Department of Energy, Division of High Energy Physics, under Grant No. DE-F605-85ER40226, and the Vanderbilt University Research Council. Discussions with, and encouragement from, V. Berezinskii, P. Biermann, D. Cline, G. Gelmini, the late D. Schramm, T. Tejima, Y. Takahashi, and A. Watson, and a critical reading by S. Dipthe Wick, are acknowledged and appreciated.